\documentstyle[aps,eqsecnum,preprint,tighten,amsmath]{revtex}
\begin{document}
\title{Finite-size scaling in systems with long-range interaction}
\author{Hassan Chamati}
\address{
Institut f\"ur Theoretische Physik, Technische Hochshule
Aachen, 52056 Aachen, Germany\\
and\\
Institute of Solid State Physics, 72 Tzarigradsko Chauss\'ee,
1784 Sofia, Bulgaria}
\date{September 3, 2001}
\maketitle
\draft
\begin{abstract}
The finite-size critical properties of the ${\cal O}(n)$ vector
$\varphi^4$ model, with long-range interaction decaying
algebraically with the interparticle distance $r$ like
$r^{-d-\sigma}$, are investigated. The system is confined to a
finite geometry subject to periodic boundary condition. Special
attention is paid to the finite-size correction to the bulk
susceptibility above the critical temperature $T_c$. We show that
this correction has a power-law nature in the case of pure
long-range interaction i.e. $0<\sigma<2$ and it turns out to be
exponential in case of short-range interaction i.e. $\sigma=2$.
The results are valid for arbitrary dimension $d$, between the
lower ($d_<=\sigma$) and the upper ($d_>=2\sigma$) critical
dimensions.
\end{abstract}

\pacs{PACS numbers: 05.70.Jk, 64.60.Ak, 64.60.Fr}

\section{Introduction}
The critical behaviour at a second order phase transition depends
upon the dimensionality $d$, degrees of freedom $n$, symmetry of
the Hamiltonian (either in spin-space or in coordinate space) and
interaction potentials. Generally speaking, the nature of the
potential of the model under consideration can describe different
physical situations (for a review  see reference
\cite{aharony76}). The simplest interaction potential, which has
attracted the attention of investigators, is the one corresponding
to long-range ferromagnetic interaction decaying algebraically
with the spin interdistance $r$ as $r^{-d-\sigma}$, where $d$ is
the dimension of the system and $\sigma$ is the parameter
controlling the range of the interaction. The interest in such
type of interaction is tightly related to the exploration of the
critical behaviour of systems with restricted dimensionality, in
which no phase transitions occur otherwise.

The investigation of systems with long-range interaction was
initiated by Joyce in his paper on the ferromagnetic spherical
model~\cite{joyce66}. The results of Joyce were generalized to the
${\cal O}(n)$ vector $\varphi^4$ model by means of perturbation
theory in combination with the renormalization group techniques
\cite{fisher72,sak73,yamazaki77,yamazaki80,honkonen89} and the
$1/n$-expansion \cite{suzuki73}. These investigations were also
extended to dynamical critical phenomena (see Ref. \cite{chen2000}
and references therein). Computer simulations also contributed in
the exploration of the critical properties of such systems
\cite{luijten95,romano96,bayong99}. The results of these
simulations, obtained by means of the Monte Carlo method,
concerned mainly systems with classical critical behaviour, in the
sense that the critical exponents are those belonging to the
Landau theory. Rigorous results were obtained for low dimensional
systems with long-range interaction (see reference
\cite{romano2000,luijten2001} and references therein).

The analytic exploration of the scaling properties of confined
systems with long-range interaction took its starting point on the
spherical model. The reason for choosing that model is the
relatively simple nature of the mathematical expression entering
the equations characterizing its thermodynamics (for a complete
set of references on the subject see reference \cite{brankov00}).
Very recently these investigations were extended to ${\cal O}(n)$
vector models \cite{luijten99,chamati2001}, using the
renormalization group approach and the $\varepsilon$ expansion to
the one-loop order. In reference \cite{luijten99} the Binder
cumulant has been evaluated at the vicinity of the critical
temperature. It has been found that the expression for that
quantity can be deduced just by choosing an appropriate rescaling
of the parameters in that evaluated for short-range interaction
case. The authors of reference \cite{chamati2001} evaluated the
susceptibility at the critical temperature $T_c$, as well as in
the region, determined by the condition $L/\xi\gg1$, where $L$ is
the linear size of the system and $\xi$ - the bulk correlation
length. In this region the bulk critical behaviour dominates the
finite-size critical one. It has been shown that the finite-size
correction to the bulk critical properties of the system has a
power-law nature. This result is distinct from that obtained for
the case of short-range interaction, where the finite-size
correction falls-off exponentially.

In this paper we will investigate the finite-size scaling in a
system with ferromagnetic long-range interaction at a fixed
dimension $d$ with $d_<<d<d_>$. The parameters $d_<=\sigma$ and
$d_>=2\sigma$ are, respectively, the lower and upper critical
dimensions of the model, with $\sigma\leq2$. To this end we will
use the approach developed in references
\cite{dohm85,shloms89,shloms90}. This method has been used
successfully for the evaluation of critical exponents, as well as,
critical amplitudes for various thermodynamic functions in the
$\varphi^4$ model with short-range potential. The most important
property of this approach is that the quadratic
(temperature-dependent) term does not enter explicitly the
expansions, which can be used for both sides of the critical
temperature $T_c$. This method has been also used for the
investigation of the theory of finite size-scaling
\cite{esser95,chen99,chen00}. While a perfect agreement between
the analytical results of reference \cite{esser95} and those
obtained by Monte Carlo simulations has been reported in reference
\cite{chen96}, references \cite{chen99,chen00} showed
disagreements with some of the known results. In particular we
would like to emphasize their finding of the non-exponential decay
of finite-size corrections to the bulk critical behaviour.

We will consider spin system consisting of $n$-component unit
vectors associated with $d$-dimensional lattice and interacting
via a pair potential of the general translationally invariant
form. The Hamiltonian of this model is given by
\begin{equation}\label{model}
\beta{\cal H}\left\{\varphi\right\}=\frac12\int_V d^d{\bbox x}
\left[a\left(\nabla\varphi\right)^2+
b\left(\nabla^{\sigma/2}\varphi\right)^2+r_0\varphi^2
+\frac12u_0\varphi^4\right],
\end{equation}
where $\varphi$ is a short hand notation for the space dependent
$n$-component field $\varphi({\bbox x})$, $r_0=r_{0c}+t_0$
($t_0\propto T-T_c$), $a$, $b$ and $u_0$ are model constants.
$V\equiv L^d$ is the volume of the system. In equation
(\ref{model}), we assumed $\hbar=k_B=1$ and the size scale is
measured in units in which the velocity of excitations $c=1$. We
note that the operator $\nabla^\sigma$ is defined by its Fourier
transform
$$
\nabla^\sigma f({\bbox x})\equiv-L^{-d}\sum_{\bbox k}
\int{d^d\bbox x'}e^{i{\bbox k}({\bbox x-\bbox x'})}|\bbox k|^\sigma
f({\bbox x}').
$$
The parameter $\beta$ is set for the inverse temperature. Here we
will consider periodic boundary conditions. This means
\begin{equation}
\varphi(x)=L^{-d}\sum_{\bbox k}\varphi({\bbox k})
\exp\left(i{\bbox k}\cdot {\bbox x}\right),
\end{equation}
where ${\bbox k}$ is a discrete vector with components $k_i=2\pi
n_i/L$ $\left(n_i=0,\pm1,\pm2,\cdots,\ i=1,\cdots,d\right)$ and a
cutoff $\Lambda\sim a^{-1}$ ($a$ is the lattice spacing). In this
paper, we are interested in the continuum limit {\it i.e.} $a\to0$. As
long as the system is finite we have to take into account the
following assumptions $L/a\to\infty$, $\xi\to\infty$ while $\xi/L$
is finite.

The critical behaviour of the model Hamiltonian (\ref{model}), has
been investigated in details in the early 70's. The main focus of
interest has been turned to the evaluation of the critical
exponents. Using renormalization group techniques, it has been
shown that the critical behaviour of this model is dominated by
the long-range interaction for $0<\sigma<2$
\cite{fisher72,yamazaki77}. In this case the critical exponents
are $\sigma$ dependent, in particular the Fisher exponent
$\eta_\sigma=2-\sigma$. As long as $\sigma$ becomes of the same
order as $2-\eta_2$, where $\eta_2$ is the Fisher exponent of the
short-range model, a crossover from the long-range critical
behaviour to the short-range critical one takes place
\cite{sak73,yamazaki80,honkonen89}. For $\sigma\geq2-\eta_2$ the
critical behaviour is dominated by the short-range interaction and
the critical exponents are those of the pure short-range model.
Another way to establish the relevance of the long range term for
$\sigma<2-\eta_2$ is presented in reference \cite{cardy96}.

The plan of this paper is as follows. In Section~\ref{bulksystem}
we describe the renormalization scheme for the bulk $\varphi^4$
theory with long-range interaction. In Section \ref{finitesystem}
we discuss the effects of confined geometries on the bulk critical
behaviour. We investigate the influence of the long-range
interaction on the finite-size correction to the bulk critical
behaviour. In Section~\ref{discussion} we discuss our results
briefly. An appendix is added in order to complement the results
of Section~\ref{finitesystem}.

\section{The bulk system}\label{bulksystem}
\subsection{Bare theory}
For simplicity here we will consider the model with pure
long-range ferromagnetic interaction i.e. the parameter $\sigma$
controlling the range of the interaction is smaller than $2$,
value characterizing the short range interaction. In other words, we
will consider the model (\ref{model}) with the parameters $a=0$
and $b=1$. We believe that one has first to understand this model
before starting to explore the model with both long and
short-range interactions present. The investigation of the
critical properties of the complex structure of this model will
presented elsewhere.

It has been shown that the Hamiltonian with pure
long-range interaction can be treated with field theoretical
renormalization group techniques \cite{yamazaki77}. The
renormalization constants as well as the field theoretic functions
were calculated. A superficial discontinuity of the anomalous
dimension of field theories occurs as soon as $\sigma=2$. This
however is true only at this particular point, and as long as we
are far from that point, we can use the model with the spectrum
$r_0+|{\bbox k}|^\sigma$. In this case the renormalization
constants, to one loop order, are given by:
\begin{mathletters}\label{renormalization}
\begin{eqnarray}
Z_\varphi&=&1+{\cal O}(u^2)\\
Z_r&=&1+\frac{n+2}\varepsilon u+{\cal O}(u^2)\\
Z_u&=&1+\frac{n+8}\varepsilon u+{\cal O}(u^2).
\end{eqnarray}
\end{mathletters}
Here, as usual, $Z_\varphi$ is the scaling field amplitude, $Z_u$
the coupling constant renormalization, and $Z_r$ the
renormalization of the $\varphi^2$ insertions in the critical
theory. The parameter $\varepsilon=2\sigma-d$ denotes the
deviation from the upper critical dimension.

The application of the method proposed by the authors of
references \cite{dohm85,shloms89,shloms90} to systems with pure
long-range interaction can be established easily following the way
this has been done in combination with the $\varepsilon$ expansion
\cite{yamazaki77}. In particular we turn our attention to the
inverse bare susceptibility, related to the two-point vertex
function, at finite external wave-number $\bbox k$
\begin{equation}
\frac1{\overset{0}{\chi}(\bbox k)}=\overset{0\ \ \ }{\Gamma^{(2)}}(\bbox
k,r_0,u_0,\Lambda,d).
\end{equation}
This function is, of course, well defined at $T\gtrsim T_c$ for
dimensions $d>\sigma$. The parameter $r_0$ is, as usual, a linear
function of the reduced temperature $t=(T-T_c)/T_c$. Its critical
value is determined by the condition $\overset{0}{\chi}(\bbox
0)^{-1}=0$, from which we obtain a natural implicit definition for
the function
\begin{equation}\label{r0c}
r_{0c}\equiv r_{0c}(u_0,\Lambda,d).
\end{equation}
Following reference \cite{dohm85}, we will consider the difference
$r_0-r_{0c}$, which is a function of the correlation length (i.e.
$r_0-r_{0c}=h(\xi,u_0,d)$), instead of $r_0$ itself into the
expressions for the  different vertex functions. So, we will
consider the vertex functions $\overset{0\ \ \ }{\Gamma^{(N)}}$ as
depending on the parameters $r_0-r_{0c}$, $u_0$, $\Lambda$ and
$d$. We will denote the dimensionally regularized vertex function
by $\overset{0\ \ \ }{\Gamma^{(N)}}\equiv\overset{0\ \ \
}{\Gamma^{(N)}} \left(r_0-r_{0c},u_0,d\right)$. Correspondingly
the dimensionally regularized critical parameter $r_{0c}$ will be
denoted by $r_{0c}(u_0,d)$. From simple dimensional arguments, one
obtains the relation
\begin{equation}\label{ru0}
r_{0c}\propto u_0^{\sigma/\varepsilon}.
\end{equation}
The method used here differs from the ones when the $\varepsilon$
expansion come into play, by the fact that the critical parameter
$r_{0c}$ is
non-vanishing within the dimensional regularized theory. It is
clear that the use of the $\varepsilon$ expansion implies an
apparent vanishing of $r_{0c}$, as seen formally from the relation
(\ref{ru0}) between $r_{0c}$ and $u_0$, which for infinitesimal
$\varepsilon$ does not yield a contribution at finite order to the
perturbation theory.

In order to make clear the definitions introduced along this
section, we will present here the one-loop results. For the two
point vertex function we get
\begin{equation}
\overset{0\ \ \ }{\Gamma^{(2)}}=r_0+{\bbox
k}^\sigma-(n+2)u_0A_{d,\sigma}
\frac{r_0^{-\varepsilon/\sigma}}\varepsilon+{\cal O}(u_0^2),
\end{equation}
where the geometrical factor $A_{d,\sigma}$ is defined by
\begin{equation}
A_{d,\sigma}=\frac2{(4\pi)^{d/2}\Gamma(d/2)}
\Gamma\left(1+\frac\varepsilon\sigma\right)
\Gamma\left(1-\frac\varepsilon\sigma\right).
\end{equation}
The four point vertex function is
\begin{equation}
\overset{0\ \ \ }{\Gamma^{(4)}}=u_0-(n+8)u_0^2
A_{d,\sigma}\left(1-\frac\varepsilon\sigma\right)
\frac{r_0^{-\varepsilon/\sigma}}\varepsilon+{\cal O}(u_0^2).
\end{equation}
In the next subsection we will discuss the renormalization of the
bare theory for fixed dimension $d$ confined between the lower and
the upper critical dimensions given by $\sigma$ and $2\sigma$,
respectively.

\subsection{Renormalized theory}
It is well known, that the perturbative results of the bare theory
do  not provide a correct description in the critical region
$\xi\to\infty$, for dimensions below the upper critical dimension.
This can be outwitted by taking advantage of the ideas of the
renormalized theory, which furnishes a mapping from the critical
region to non-critical one. Our starting point is the expression
for the $N$-point vertex functions $\tilde\Gamma_0^N(\xi,u_0,d)$
obtained from  $\overset{0\ \ \ }{\Gamma^{(N)}}
\left(r_0-r_{0c},u_0,d\right)$ by  switching from the variable
$r_0-r_{0c}$ to $\xi$. This is possible  because of the fact that
the reduced temperature is tightly related to the
correlation length. The deviation of the parameter $r_0$ from its
critical value will be a fixed quantity and hence the correlation
length will have the same property.

We treat the theory by using the minimal subtraction scheme at
fixed dimension $d$. To this end, we introduce the renormalized
quantities
\begin{mathletters}\label{renorm}
\begin{eqnarray}
\varphi(x)&=&Z_\varphi^{1/2}\varphi_R(x)\\
r_0&=&r_{0c}+Z_rr, \ \ \ r_0-r_{0c}>0 \\
A_{d,\sigma}u_0&=&\mu^\varepsilon Z_uZ_\varphi^{-2}u.
\end{eqnarray}
\end{mathletters}
Then the renormalized vertex functions takes the form
\begin{equation}\label{gammatilde}
\tilde\Gamma^{(N)}(\xi,u,\mu,d)=Z_\varphi^{N/2}
\tilde\Gamma_0^N(\xi,\mu^\varepsilon Z_uZ_\varphi^{-2}u
A_{d,\sigma}^{-1},d).
\end{equation}
In addition, we require that the renormalization constants $Z_\varphi$
and $Z_u$
absorb just the poles of $\tilde\Gamma_0^N$ at the upper critical
dimension, which turns out to be $2\sigma$ for the model under
consideration. In equations (\ref{renorm}) and (\ref{gammatilde})
the parameter $\mu$ is an inverse reference length, which will be
chosen as the amplitude of the asymptotic bulk correlation length.

In the following we proceed in a standard way, by deriving a
differential renormalization-group equation for the vertex
function $\tilde\Gamma^{(N)}$. This is achieved by taking
the derivative of equation
(\ref{gammatilde}) with respect to $\mu$ at fixed constant $u_0$
and $r_0-r_{0c}$. This leads to
\begin{equation}\label{RGequation}
[\mu\partial\mu+\beta_u(u,\varepsilon)\partial_u+
\frac12N\zeta_\varphi(u)]
\tilde\Gamma^{(N)}(\xi,u,\mu,d)=0,
\end{equation}
with
\begin{equation}\label{RGfunctions}
\beta_u(u,\varepsilon)=(\mu\partial_uu)_0, \ \ \ \ \ \ \ \ \
\zeta_\varphi(u)=(\mu\partial_\mu\ln Z_\varphi^{-1})_0.
\end{equation}
Here the subscript $0$ indicates that the differentiation is
performed at fixed parameters of the bare theory. Using the method
of characteristics, a formal solution of the renormalization-group
differential equation (\ref{RGequation}) is given by
\begin{equation}\label{formal}
\tilde\Gamma^{(N)}(\xi,u,\mu,d)=\tilde\Gamma^{(N)}(\xi,u(\ell),\ell\mu,d)
\exp\left(\frac N2\int_1^\ell\zeta_\varphi(\ell')\frac{d\ell'}{\ell'}\right).
\end{equation}
Here $\zeta(\ell)(u(\ell),d)$ and $u(\ell)$ is the solution of
flow equation
\begin{equation}
\ell\frac{d u(\ell)}{d\ell}=\beta_u[u(\ell),d]
\end{equation}
with the initial condition $u(\ell)=u$. The most convenient choice
for the flow parameter $\ell$ is
\begin{equation}\label{ellmu}
\ell\mu=\xi^{-1}.
\end{equation}
The renormalized vertex functions can be written as
\begin{equation}\label{firstf}
\tilde\Gamma^{(N)}(\xi,u,\mu,d)=\xi^{-d+(d-\sigma)\frac N2}
f^{(N)}(\mu\xi,u,d),
\end{equation}
where the amplitude functions $f^{(N)}$ are dimensionless. The
renormalizability of the $\varphi^4$ model with long-range
interaction for dimensions $d$ less than the upper critical
dimension $2\sigma$ is a warranty for the finiteness of the
$\tilde\Gamma^{(N)}$ at fixed $u$, $\mu$ and $\xi$. From
(\ref{formal}), (\ref{ellmu}) and (\ref{firstf}), we obtain the
expression
\begin{equation}\label{ffinal}
f^{(N)}(\mu\xi,u,d)=f^{(N)}(1,\mu(\ell),d)
\exp\left(\frac
N2\int_1^\ell\zeta_\varphi(\ell')\frac{d\ell'}{\ell'}\right).
\end{equation}
As usual, equation (\ref{ffinal}) lies in the basis of the mapping
of the amplitude function $f^{(N)}(\mu\xi,u,d)$ from the critical
region, where the perturbation theory breaks
down to the noncritical region, where the perturbation theory is
applicable.

\subsection{Asymptotic regime}
In the asymptotic limit, determined by ($\ell\to0$,
$\xi\to\infty$), the coupling $u(\ell)$ approaches the fixed point
$u^*=u(0)$, which is the zero of the $\beta$ function of the
theory i.e.
\begin{equation}
\beta_u(u^*,0,d)=0.
\end{equation}
For $\xi\to\infty$, equation (\ref{ffinal}), takes the asymptotic form
\begin{equation}
f^{(N)}(\mu\xi,u,d)\sim A^{(N)}f^{(N)}(1,u^*,0,d)(\mu\xi)^{N\eta/2},
\end{equation}
with the critical exponent $\eta=-\zeta_\varphi(u^*,0,d)$ and the
nonuniversal amplitude
\begin{equation}
A^{(N)}=\exp\left(\frac N2\int_1^0[\zeta_\varphi(\ell')-
\zeta_\varphi(0)]\frac{d\ell'}{\ell'}\right).
\end{equation}

The results obtained here can be applied very easily to the
particular case of the bare susceptibility corresponding to the
particular value $N=2$. This is given by
\begin{equation}
\chi=Z_\varphi(u,d)\xi^2[f^{(2)}(1,u(\ell),d)]^{-1}
\exp\left(\int_\ell^0\zeta_\varphi(\ell')\frac{d\ell'}{\ell'}\right).
\end{equation}
The critical behaviour above $T_c$ reads
\begin{equation}
\chi={\cal A}^+_\chi t^{-\gamma},
\end{equation}
where
\begin{equation}
{\cal A}^+_\chi=\xi_0^{2}Z_\varphi(u,d)
[A^{(2)}f^{(2)}(1,u^*,0,d)]^{-1}.
\end{equation}
Here we have used $\mu=\xi^{-1}_0$ and the asymptotic form
$\xi=\xi_0t^{-\nu}$ with $\gamma=\nu(2-\eta)$.

\section{The finite system}\label{finitesystem}
\subsection{The effective Hamiltonian}
For the evaluation of the effective Hamiltonian of the finite
system, we need to calculate the free energy per unit volume. This
is defined by
\begin{equation}\label{free_energy}
f=-L^{-d}\ln{\cal Z}; \ \ \ \ \
{\cal Z}=\int{\cal D}\varphi\exp(-\beta{\cal H}).
\end{equation}
Here we are also interested in the expression of the
susceptibility, which is defined by
\begin{equation}\label{chi}
\chi=\int d^d{\bbox x}\left<\varphi({\bbox x})\varphi(0)\right>
=\frac1{\cal Z}\int d^d{\bbox x}\int{\cal D}
\varphi\varphi({\bbox x})\varphi(0)\exp(-\beta{\cal H}).
\end{equation}

Following references \cite{brezin85,rudnick85}, we split the field
$\varphi=\Phi+\Sigma$ into a mode independent part
\begin{equation}\label{zeromode}
\Phi=L^{-d}\int d^d{\bbox x}\varphi({\bbox x}),
\end{equation}
which is equivalent to the magnetization and a part depending upon
the non zero modes
\begin{equation}\label{nozeromode}
\Sigma=L^{-d}{\sum_{\bbox k}}'\varphi({\bbox k})\exp(i{\bbox k}\cdot
{\bbox x}).
\end{equation}
Consequently (\ref{model}) is decomposed as
\begin{equation}
{\cal H}={\cal H}_0+{\cal H}_I(\Phi,\Sigma),
\end{equation}
with the zero-mode Hamiltonian
\begin{equation}\label{zeromodeh}
{\cal H}_0=\frac12L^d\left(r_0\Phi^2+\frac12u_0\Phi^4\right).
\end{equation}
Whence the partition function takes the form
\begin{equation}\label{h0+g0}
{\cal Z}=\int_{-\infty}^\infty d\Phi\exp[-({\cal H}_0+
\overset{0}{\Gamma}(\Phi^2))],
\end{equation}
where
\begin{equation}
\overset{0}{\Gamma}(\Phi^2)=-\ln\int{\cal D}\Sigma\exp[-{\cal H}_I]
\end{equation}
contains the contribution from higher order.

Now instead of the decomposition of reference \cite{rudnick85} we
will use the modified perturbation theory proposed in reference
\cite{esser95}. There, an appropriate decomposition of the ${\cal
O}(n)$ vector $\varphi^4$ model with short-range interaction has
been presented. This way has been proven to give a good
quantitative results above, as well as below the critical
temperature $T_c$. Applying that method to our model we find, that
the higher-mode dependent part ${\cal H}_I$ can be split into
\begin{eqnarray}
{\cal H}_I&=&\frac12\int d^dx[(r_0+(n+2)u_0M^2_0)\Sigma^2+
(\nabla^{\sigma/2}\Sigma)^2]\nonumber\\
&&+\frac12\int
d^dx[3u_0(\Phi^2-M_0^2)\Sigma^2+2u_0\Phi\Sigma^3+\frac12u_0\Sigma^4],
\end{eqnarray}
where we have introduced the magnetization
\begin{equation}
M_0^2=\left<\phi^2\right>.
\end{equation}
Further a diagrammatic expansion of $\overset{0}{\Gamma}(\Phi^2)$
can be represented by two point vertices proportional to
$u_0(\Phi^2-M_0^2)$ in addition to the three and four point
vertices $\sim \ u_0\Phi$ and $\sim \ u_0$. In this way, the
finite-size perturbation theory is obtained as an expansion of
$\overset{0}{\Gamma}(\Phi^2)$ in powers of $u_0(\Phi^2-M_0^2)$,
$u_0$ and $u_0^2\Phi^2$. In the following, instead of
$\overset{0}{\Gamma}(\Phi^2)$
we consider $L^{-d}\overset{0}{\Gamma}(\Phi^2)$ which remains
finite in the limit $L\to\infty$. To the leading order the
finite-size correction reads
\begin{eqnarray}\label{gamma0}
L^{-d}\overset{0}{\Gamma}(\Phi^2)&=&\frac12{\sum_{\bbox k}}'
\ln(r_0+(n+2)u_0M^2_0+{\bbox
k}^\sigma)+\frac{n+2}2u_0\Phi^2S_1(r_0)
\nonumber\\
&&-\frac{n+8}4u_0^2(\Phi^4+2M_0^2\Phi^2)S_2(r_0)+{\cal O}
(u_0,u_0^2\phi^2,u_0^3(\Phi^2-M_0^2)^3),
\end{eqnarray}
where
\begin{equation}
S_m(r)=L^{-d}{\sum_{\bbox k}}'(r+(n+2)u_0M^2_0 +{\bbox
k}^\sigma)^{-m}.
\end{equation}

Here we will investigate only the high temperature regime i.e.
$T\gtrsim T_c$, so we will set $M_0^2=0$. A rearrangement of
equations (\ref{zeromodeh}) and (\ref{gamma0}) leads to the
effective Hamiltonian
\begin{equation}\label{effHam}
{\cal H}_{\rm eff}(r_0,u_0,L,\Phi^2)=\frac12L^d
[R\Phi^2+\frac12U\Phi^4+{\cal O}(\Phi^6)],
\end{equation}
where
\begin{mathletters}\label{effconst}
\begin{eqnarray}
R&=& r_0+(n+2)u_0L^{-d}{\sum_{\bbox k}}'(r_0+{\bbox k}^\sigma)^{-1},\\
U&=& u_0-(n+8)u_0^2L^{-d}{\sum_{\bbox k}}'(r_0+{\bbox
k}^\sigma)^{-2}.
\end{eqnarray}
\end{mathletters}
Finally, the free energy (\ref{free_energy}) reads
\begin{mathletters}
\begin{eqnarray}
f&=&L^{-d}\overset{0}{\Gamma}(0)-L^{-d}\ln{\cal Z}_{\rm eff},\\
{\cal Z}_{\rm eff}&=&\int_{-\infty}^\infty d\Phi
\exp[{\cal H}_{\rm eff}(r_0,u_0,L,\Phi^2)]
\end{eqnarray}
\end{mathletters}
and averages such as $\left<\Phi^{2p}\right>$ are calculated from
(\ref{chi}) as
\begin{equation}\label{momenta}
\left<\Phi^{2p}\right>=\frac1{\cal
Z}_{\rm eff}\int_{-\infty}^\infty d\Phi\Phi^{2p}
\exp(-{\cal H}_{\rm eff}(r_0,u_0,L,\Phi^2)).
\end{equation}
Let us note that for the purpose of comparison with numerical
results it is convenient to keep the exponential structure of the
integrand in the various thermodynamic functions. Notice that an
appropriate rescaling of the field $\Phi$ in equation
(\ref{momenta}), permits to write the momenta
$\left<\Phi^{2p}\right>$ as a function the `scaling variable'
\begin{equation}\label{zsca}
z=RL^{d/2}U^{-1/2}.
\end{equation}
Whence
\begin{equation}
\left<\Phi^{2p}\right>=(u_0L^{d})^{-p/2}\frac{\int_0^\infty
dxx^{n+2p-1}\exp\left(-\frac12zx^2-\frac14x^4\right)}
{\int_0^\infty dxx^{n-1}\exp\left(-\frac12zx^2-\frac14x^4\right)}.
\end{equation}
These functions can be expressed in terms of the confluent
hypergeometric function (see Appendix \ref{app1}).

\subsection{Renormalization}
As it has been mentioned in the previous section, in this paper,
we will use the approach of references
\cite{dohm85,shloms89,shloms90}. Here we will consider the limit
of an infinite cutoff i.e. $\Lambda\to\infty$ at fixed
$r_0-r_{0c}$, where $r_{0c}$ is the bulk critical value of $r_0$.
Let us recall that the advantage of using this method is its
direct application to fixed space dimensionality $d$, without
resorting to the $\varepsilon$ expansion. For systems with
short-range interaction this method has been tested and accurate
results, for various thermodynamic functions has been, obtained
\cite{esser95}. Since $r_{0c}$ is of order ${\cal
O}(u_0^{\sigma/2})$ in dimensionally regularized form, we find it
safe to replace in our expressions $r_0$ by the difference
$r_0-r_{0c}$. The effective Hamiltonian will be function of
$r_0-r_{0c}$ but not of $r_0$ itself. The effective constants $R$
and $U$ defined in (\ref{effconst}) are also functions of this
deviation, which itself is a function of the correlation length.
The perturbation approach consists of an expansion of ${\cal
H}_{\rm eff}$ with respect to the renormalized counterpart $u$ of
the coupling constant $u_0$ at fixed dimension. Since the finite
nature of the geometry of the system does not alter the
ultraviolet divergences, the usual bulk renormalizations are
enough to describe the scaling properties the finite system.

Here we are interested in the evaluation of the explicit
expressions of various thermodynamic quantities for the finite
system. Different crossover regions, in terms of renormalized
parameters, are under consideration. The corresponding
renormalized effective coupling constants to equations
(\ref{effconst}) take the form
\begin{mathletters}\label{RUfinal}
\begin{eqnarray}
R&=&r(\ell)+(n+2)u(\ell)(\mu\ell)^\sigma
\left[\frac{r(\ell)(\mu\ell)^{-\sigma}}\varepsilon
\left[1-\left(\frac{r(\ell)}{(\mu
\ell)^\sigma}\right)^{-\varepsilon/\sigma}\right]\right.\nonumber\\
& &\left.+\frac{(\mu L)^{\varepsilon-\sigma}}{A_{d,\sigma}}
F_{d,\sigma}(r(\ell)L^\sigma)\right],\\
U&=&u(\ell)+u^2(\ell)(n+8)\left[\frac1\varepsilon\left[1-\left(\frac{r(
\ell)}{(\mu\ell)^\sigma}\right)^{-\varepsilon/\sigma}\right]+\frac1\sigma
\left(\frac{r(\ell)}{(\mu
\ell)^\sigma}\right)^{-\varepsilon/\sigma}\right.\nonumber\\
&&\left.+\frac{(\mu L)^{\varepsilon}}{A_{d,\sigma}}
F_{d,\sigma}'(r(\ell)L^\sigma)\right],
\end{eqnarray}
\end{mathletters}
where we have used the definitions \cite{chamati2001}
\begin{mathletters}
\begin{eqnarray}
S_1(r)&=&-A_{d,\sigma}\varepsilon^{-1}r^{1-\varepsilon/\sigma}
+ L^{\sigma-d}F_{d,\sigma}(rL^\sigma).\\
S_2(r)&=&A_{d,\sigma}\varepsilon^{-1}
\left(1-\frac\varepsilon\sigma\right)
r^{-\varepsilon/\sigma}- L^{\sigma-d}\frac\partial{\partial r}
F_{d,\sigma}(rL^\sigma),
\end{eqnarray}
\end{mathletters}
and
\begin{equation}\label{It}
F_{d,\sigma}\left(y\right)=\frac1{(2\pi)^\sigma}\int_0^\infty dx
x^{\frac\sigma2-1}E_{\frac\sigma2,\frac\sigma2}
\left(-\frac{yx^{\sigma/2}}{(2\pi)^\sigma}\right)
\left[\left(\sum_{\ell=-\infty}^\infty e^{-x\ell^2}\right)^d
-1-\left(\frac{\pi}{x}\right)^{d/2}\right].
\end{equation}
Here the function
\begin{equation}\label{mittag}
E_{\alpha,\beta}(z)=\sum_{k=0}^\infty\frac{z^k}
{\Gamma\left(\alpha k+\beta\right)}
\end{equation}
is the so called Mittag-Leffler type functions. For a more recent
review on these functions and others related to them, and their
application in statistical and continuum mechanics see
reference~\cite{mainardi97}. A brief summary of some of the
properties of the Mittag-Leffler functions are presented in
reference \cite{chamati2001}.

By making the choice such that the flow parameter $\ell=\ell(t,L)$
satisfies the following relation
\begin{equation}\label{ell}
\xi^{-\sigma}+L^{-\sigma}=\mu^\sigma\ell^\sigma,
\end{equation}
in the critical region, equations (\ref{RUfinal}) imply the
scaling forms
\begin{equation}\label{elltL}
R(\ell)=\mu^\sigma\ell^\sigma\tilde R(tL^{1/\nu}), \ \ \ \ \ \
U(\ell)=\tilde U(tL^{1/\nu}).
\end{equation}
Using the asymptotic expressions for $R(\ell)$ and $U(\ell)$, we
obtain the asymptotic form of the variable $z$, defined in
(\ref{zsca}). It is given by
\begin{equation}
z(tL^{1/\nu})=R(\ell)\mu^{-2}\ell^{-2}(L\mu\ell)^{d/2}
A_d^{1/2}[U(\ell)]^{-1/2}.
\end{equation}
From this expression, one can convince himself, easily, that the
function $z(tL^{1/\nu})$ has an expansion in terms of $\sqrt{u^*}$.

\subsection{The Susceptibility}
In this section, we present our result for the finite-size scaling
function of the magnetic susceptibility. Here we give the general
expression, obtained by expanding with respect to the coupling
constant $u$ about its fixed point $u^*$. The dimension is kept
fixed i.e. we are not going to expand in the vicinity of the upper
critical dimension as it has been done earlier \cite{chamati2001}.
Furthermore, we won't expand the exponential weight of the
integrand in equation (\ref{momenta}). In this way consistency
with the correct one-loop bulk expressions is ensured in the bulk
limit, reached by sending the size $L$ of the system to infinity.
As it has been mentioned in the previous sections the arbitrary
reference length $\mu^{-1}$ of the renormalized theory will be
chosen as the amplitude of the correlation length i.e.
$\mu^{-1}=\xi_0$.

In reference \cite{brezin82}, it has been shown that, for periodic
boundary conditions, the ultraviolet divergences of the
susceptibility $\chi$ of finite systems in the continuum limit
i.e. $a\to0$ are identical to those of the bulk susceptibility
$\chi_b$. In reference \cite{chamati2001}, the application of this
method has been extended to systems with long-range interaction
and some results for the susceptibility in the asymptotic regimes
$L/\xi\gg1$, as well as $L/\xi\ll1$ has been obtained. The results
obtained there were limited to dimensions close to the upper
critical one. Here we will extend these results to the whole
interval of dimensions between the lower critical dimension
$d_<=\sigma$ and the upper critical one $d_>=2\sigma$.

The average $\left<\phi^2\right>$, entering the definition of the
susceptibility (\ref{chi}), is defined with the statistical weight
$\exp[-({\cal H}_0+\overset{0}{\Gamma})]$ given in equation
(\ref{h0+g0}). Obviously, in the bulk limit we recover the bulk
susceptibility i.e. $\lim_{L\to\infty}\chi=\chi_b$.

For systems confined to a finite geometry, the renormalized
susceptibility $\chi_R$, as a function of $r_0-r_{0c}$ , $u_0$,
and $L$, can be introduced as
\begin{equation}\label{chi_R}
\chi_R(\xi,u,L,\mu,d)=Z_\varphi^{-1}\chi(\xi,\mu^\varepsilon
Z_uZ_\varphi A_{d,\sigma}u,L,d),
\end{equation}
where $Z_\varphi$ and $Z_u$ are the bulk $Z$ amplitude factors
defined in Section~\ref{bulksystem}.

A renormalization-group equation for $\chi_R$ is obtained by
deriving expression (\ref{chi_R}) for the susceptibility with
respect to the parameter $\mu$ at fixed $r_0-r_{0c}$ (function of
the correlation length), $u_0$ and $d$. Since the linear size $L$
of the system does not renormalize, we get
\begin{equation}\label{finite_chi}
\left[\mu\partial_\mu+\beta_u\partial_u-\zeta_\varphi\right]
\chi_R(\xi,u,L,\mu,d)=0,
\end{equation}
where the renormalization group functions $\beta(u)$ and
$\zeta(u)$ are defined in equation (\ref{RGfunctions}) for the
bulk theory. A formal solution for this equation is given by
\begin{equation}\label{chi_Rsol}
\chi_R(\xi,u,L,\mu,d)=\chi_R(\xi,u(\ell),L,\ell\mu,d)
\exp\left(\int_\ell^1\zeta_\varphi(\ell')\frac{d\ell'}{\ell'}\right),
\end{equation}
where the parameter $\ell$ can be chosen arbitrarily. The most
convenient choice is that, for which $\ell$ satisfies the relation
(\ref{ell}). Remark that the renormalization group equations
(\ref{RGequation}) for the bulk system and (\ref{finite_chi}) for
the system confined to the finite geometry are similar. However
because of the finiteness of the size $L$ of the system a careful
consideration of equations (\ref{finite_chi}) and its solution
(\ref{chi_Rsol}) is in order. In this case, we introduce the
dimensionless amplitude function $f_\chi(z)$ according to
\begin{equation}\label{dimless}
\chi_R(L,\xi,u,\mu,d)=L^2 f_\chi(\mu L,\mu\xi,u,d).
\end{equation}
In the asymptotic regime, given by $\ell\ll1$, we obtain from the
formal solution (\ref{chi_Rsol}) of the renormalization group
equation
\begin{equation}
\chi_R(L,\xi,u,\mu,d)\sim L^2\ell^\eta\left[A^{(-2)}\right]
f_\chi(\mu\ell L,\mu\ell\xi,u^*,d).
\end{equation}
Using the fact that $\mu\ell L$ and $\mu\ell\xi$ are functions
only of the ratio $L/\xi$, which follows from equation
(\ref{ell}), we can write the bare susceptibility
$\chi=Z_\varphi\chi_R$, as well as, the renormalized susceptibility
$\chi_R$ in the following scaling form
\begin{equation}\label{suscep}
\chi_R(L,\xi,u,\mu,d)\sim L^{2-\eta}\mu^\eta\left[1+\left(\frac L\xi
\right)^\sigma\right]^{\eta/\sigma}\left[A^{(-2)}\right]
Y\left(\frac L\xi\right),
\end{equation}
where $Y(z)$ is a scaling function of its argument.

Equation (\ref{suscep}) is the final expression for the
susceptibility of the finite system. This is the complete
expression in whole $L^{-1}-\xi^{-1}$ plane. In the following we
will consider different regimes with respect to the ratio $\xi/L$.
For this purpose we investigate the ${\cal O}(n)$ vector
$\varphi^4$ model. Here we turn a special attention to the
limiting case $\xi/L\gg1$.

In the remainder of this section we will investigate the behaviour
of the susceptibility using an ordinary perturbation theory.
Accordingly, the standard one-loop expression for the
inverse susceptibility above the critical point for a system
confined to a finite geometry reads
\begin{equation}
\chi^{-1}=r_0+(n+2)u_0L^{-d}\sum_{\bbox k}(r_0+\bbox k^\sigma)^{-1}+
{\cal O}(u_0^2).
\end{equation}
The susceptibility for the bulk system is obtained by sending the
size of the system to infinity i.e. the sum in the last equation
is replaced by integrals. It is
\begin{equation}
\chi_b^{-1}=r_0+(n+2)u_0\int d^d{\bbox k}(r_0+\bbox k^\sigma)^{-1}+
{\cal O}(u_0^2).
\end{equation}
After some straightforward calculation one gets
\begin{equation}\label{chiord}
\chi(t,u_0,L,d)^{-1}=t\left[1+(n+2)u_0L^{2\sigma-d}(tL^\sigma)^{-2}
\left(1+tL^\sigma F_{d,\sigma}(tL^\sigma)\right)\right].
\end{equation}
From this expression, we see that it is not allowed to set $y=0$.
In other words to take the limit $t\to0$, while the size of the
system is kept fixed. In the opposite limit $L/\xi\gg1$, the right
hand side of equation (\ref{chiord}) is well defined and the
result gives the finite-size correction to the expression of the
bulk susceptibility. However, the behaviour of the function
$F_{d,\sigma}(y)$ is strongly dependent upon the value of the
parameter $\sigma$. Indeed, it depends upon the nature of the
interparticle interaction in the system. Here we will discuss the
influence of the interaction on the behaviour of the finite-size
correction.

The function $F_{d,\sigma}(y)$ has the following large $y$
asymptotic behaviour \cite{chamati2001}
\begin{equation}\label{largeya}
F_{d,\sigma}(y)\simeq-\frac1y+
\frac{2^\sigma\pi^{-d/2}\Gamma\left(\frac{d+\sigma}2\right)}
{y^2\Gamma(-\frac\sigma2)}{\sum_{\bbox l}}'\frac1{|\bbox
l|^{d+\sigma}}
\end{equation}
for the case $0<\sigma<2$, and
\begin{equation}\label{largeyb}
F_{d,2}(y)\simeq-\frac1y+d(2\pi)^{(1-d)/2}y^{(d-3)/4}e^{-\sqrt y}
\end{equation}
for the particular case $\sigma=2$. These results show that the
last term in equation (\ref{finalchi}) is just cancelled by the first
term in equations (\ref{largeya}) and (\ref{largeyb}).

In the case of long-range interaction $0<\sigma<2$, we obtain for
the susceptibility, after renormalizing the theory,
\begin{equation}\label{longrange}
\chi=\chi_b\left[1-u^*(n+2)2^\sigma\pi^{-d/2}\left(tL^\sigma\right)
^{-1-\frac d\sigma}\frac{\Gamma\left(\frac{d+\sigma}2\right)}
{\Gamma(-\frac\sigma2)} {\sum_{\bbox l}}' \ {\bbox
l}^{-d-\sigma}+{\cal O}({u^*}^2)\right]
\end{equation}
in agreement with the finite-size scaling hypothesis. Equation
(\ref{longrange}) shows that the finite-size scaling behaviour of
the system is dominated by the bulk critical behaviour, with small
correction in powers of $L$. First the power law fall-off of the
finite-size corrections to the bulk critical behaviour, due to
long-range nature of the interaction, was found in the framework
of the spherical model~\cite{singh89,brankov91}, which is believed
to belong to the same class of universality as the ${\cal O}(n)$
vector model with $n\to\infty$.

It should be noted that the above result~(\ref{longrange}) cannot
be continued smoothly to the case of short-range interaction
$\sigma=2$. In this particular case$F_{d,2}(y)$
(see equation (\ref{largeyb}))
falls off exponentially fast and, correspondingly, the finite-size
corrections to $\chi$ are exponentially small:
\begin{equation}\label{shortrange}
\chi=\chi_b\left[1 - u^*(n+2)d(2\pi tL^2)^{(1-d)/4} e^{-L\sqrt t}
+ {\cal O}\left({u^*}^{2}\right)\right].
\end{equation}
This result was obtained first in reference \cite{chen00}. Notice
that by expanding to the first order in $\varepsilon=2\sigma-d$ in
equation (\ref{longrange}) and (\ref{shortrange}), we recover, the
corresponding results (3.10) and (3.31) of reference
\cite{chamati2001}.

Here an important remark is in order. In appendix \ref{app1}, we
show that the results obtained for the susceptibility in this
section are identical to those obtained, using the mode expansion.
This is in disagreement with the conclusions of
reference~\cite{chen00} claiming that the mode expansion is
inadequate for the description of the finite-size scaling in the
region of the phase diagram determined by the condition
$L/\xi\gg1$. Especially for the case of short-range interaction
this result is in agreement with that obtained by means of Monte
Carlo.

\section{discussion}\label{discussion}
In this paper, we have investigated the finite-size scaling
properties in the ${\cal O}(n)$-symmetric $\varphi^4$ model with
long-range interaction potential decaying algebraically with the
interparticle distance. By means of the method, which consists of
the use of the minimal subtraction scheme applied to a fixed space
dimensionality developed in references
\cite{dohm85,shloms89,shloms90} for the bulk system and in
\cite{esser95,chen99,chen00} for the one with finite linear size,
we generalized the results of reference \cite{chamati2001}
obtained for the susceptibility by means of the
$\varepsilon=2\sigma-d$ expansion. The obtained results are in a
full agreement with those obtained in reference
\cite{chamati2001}.

Here we restricted our calculation to the critical domain
$T\gtrsim T_c$ and investigated the behaviour of the
susceptibility to the one-loop order in the coupling constant
$u_0$ of the bare theory. We have turned our attention to the
special case, in the phase diagram, where the bulk critical
behaviour is dominating the finite-size scaling one. We have found
that the finite-size correction falls off algebraically
(\ref{longrange}) in the case when $\sigma<2$ and exponentially
(\ref{shortrange}) in the particular case $\sigma=2$,
characterizing the short-range interaction. These results are
obtained by using two different methods: (i) The standard mode
expansion of references \cite{brezin85,rudnick85} and (ii) The
ordinary perturbation theory used in references
\cite{chen99,chen00}. We would like to mention here that our
results does not agree with those of reference \cite{chen00}
claiming that the method built on the mode expansion does not
adequately describe the critical behaviour of the finite system.

Note that by evaluating the Binder's cumulant ratio, we obtain a
result, which have the same behaviour as a function of
$\varepsilon$ as that of the result of references
\cite{luijten99,chamati2001}. It seems that to the first loop
order, the method used here does not ameliorate the result
obtained by means of the $\varepsilon$-expansion. It is possible
that higher loop order can improve the result in comparison to the
Monte Carlo method of reference \cite{luijten99}.

In this paper, we concentrated our attention to a field
theoretical model in the continuum (scaling) limit. Consequently
the cutoff is send to infinity. By accounting the finite cutoff
effects we expect that the finite-size scaling will be violated in
a similar way to the case short-range interaction \cite{chen99}.
Note that the same effects can be obtained if one takes the model
(\ref{model}) with the parameter $\sigma$ controlling the range of
the interaction to be larger than the value, $\sigma=2$,
characterizing the short potential. This would verify whether the
results obtained in references \cite{danchev2001} at the spherical
limit remains true for finite $n$.

Another possible extension of the results obtained here, in the
static limit, is the application of the finite-size scaling theory
to systems including critical dynamics. This can have direct
implication to so systems exhibiting quantum critical behaviour.

\acknowledgments

The author would like to thank Profs. V. Dohm and N.S. Tonchev for
stimulating discussions. The hospitality of the Institute of
Theoretical Physics of RWTH-Aachen is acknowledged.

\appendix
\section{Finite-size corrections to the bulk susceptibility}
\label{app1}

In the appendix, we evaluate the finite-size correction to the
bulk susceptibility in the region $L/\xi\gg1$ for arbitrary
dimension of the system. Here we use the mode expansion and we
will work in the one loop order as explained in Section
\ref{finitesystem}. We are interested in particular in dimensions
$d$, such that $\sigma<d<2\sigma$. In the leading order of the
non-zero ($\bbox k\neq\bbox0$) modes the effective Hamiltonian
(\ref{effHam}) reads
\begin{equation}\label{effHamA}
{\cal H}_{\rm eff}(r_0,u_0,L,\Phi^2)=\frac12L^d
[R\Phi^2+\frac12U\Phi^4+{\cal O}(\Phi^6)],
\end{equation}
where
\begin{mathletters}\label{effconstA}
\begin{eqnarray}
R&=& r_0-r_{0c}+(n+2)u_0\left[
L^{-d}{\sum_{\bbox k}}'(r_0-r_{0c}+{\bbox k}^\sigma)^{-1}
-\int_{\bbox k}{\bbox k}^{-\sigma}\right],\\
U&=& u_0-(n+8)u_0^2L^{-d}{\sum_{\bbox k}}'(r_0-r_{0c}+{\bbox
k}^\sigma)^{-2}.
\end{eqnarray}
\end{mathletters}
We have incorporated here the finite shift $r_{0c}=-(n+2)u_0
\int_{\bbox k}{\bbox k}^{-\sigma}+{\cal O}(u_0)$ of the parameter
$r_0$ at one loop-order. Let us recall that in general one does a
double expansion: one in the modes and the other in the coupling
constant $u_0$. In other words the fact that we are working in
fixed space dimension does not mean that the parameter $u_0$
should be kept fixed or we are not allowed to expand with respect
to the small parameter $u_0$. This observation will be of a great
importance in the following.

In the one-loop order the difference $r_0-r_{0c}$ is proportional
to the reduced temperature $t$ i.e. $r_0-r_{0c}=a t$. In the
following we will choose the coefficient $a=1$. In terms of $t$,
equations (\ref{effconstA}) reads
\begin{mathletters}\label{effconstA1}
\begin{eqnarray}
R&=&t-(n+2)u_0\Delta_1(t),\\
U&=& u_0-(n+8)u_0^2\int_{\bbox k}(t+{\bbox
k}^\sigma)^{-2}+(n+8)u_0^2\Delta_2(t),
\end{eqnarray}
\end{mathletters}
where
\begin{equation}
\Delta_m=\int_{\bbox k}(t+{\bbox k}^\sigma)^{-m}-
L^{-d}{\sum_{\bbox k}}'(t+{\bbox k}^\sigma)^{-m}.
\end{equation}

Now we will evaluate the susceptibility for the system confined to
a finite geometry. In the present approximation it is
\begin{equation}\label{chidefa}
\chi=\frac1n\sqrt\frac{L^d}{U}{\cal
G}_\chi\left(R\sqrt\frac{L^d}{U}\right),
\end{equation}
with
\begin{equation}\label{integrals}
{\cal G}_\chi(z)=\frac{\int_0^\infty
dxx^{n+1}\exp\left(-\frac12zx^2-\frac14x^4\right)}
{\int_0^\infty dxx^{n-1}\exp\left(-\frac12zx^2-\frac14x^4\right)}.
\end{equation}

The integrals appearing in equation (\ref{integrals}) can be
expressed in terms of the confluent hypergeometric function
$U(a,b;z)$ according to
\begin{equation}
\int_0^\infty dy y^{\nu-1}e^{-z\frac{y^2}2-\frac{y^4}4}=
\frac12\Gamma\left(\frac\nu2\right)U\left(\frac\nu4,
\frac12;\frac{z^2}4\right), \ \ \ \ \ \text{for} \ z\geq0.
\end{equation}
This function has a well known analytic properties (see reference
\cite{abramovitz64}).

In the region $tL^{1/\nu}\gg1$, corresponding to $z\gg1$, using
the asymptotic form of the function ${\cal G}(z)$ for large
argument, which follows from that of the confluent Hypergeometric
function, we obtain for the susceptibility
\begin{equation}\label{chiinta}
\chi=\frac1R\left[1-(n+2)\frac U{L^dR^2}\right].
\end{equation}
Here we would like to mention that this result is not the final
expression for the susceptibility of the finite system. Here it
easy to make a mistake by taking, to the lowest mode
approximation, the coupling constants $R$ and $U$ to be equal to
the initial coupling constants $r_0-r_{0c}$ and $u_0$,
respectively. One has to keep in mind that apart from the lowest
mode approximation, a loop expansion in $u_0$ is involved in the
calculations. So, here one must take into account the full
expressions of the constants $R$ and $U$ up to the lowest
(one-loop) order in $u_0$. Consequently by expanding in equation
(\ref{chiinta}) to order ${\cal O}(u_0^{2})$ we arrive to the
final result for the susceptibility
\begin{equation}\label{finalchi}
\chi=t^{-1}\left[1-(n+2)u_0\frac{L^{2\sigma-d}}{y^2}
\left(1+yF_{d,\sigma}(y)\right)\right].
\end{equation}
Here we have used $y=tL^\sigma$ and
$\Delta_1(t)\equiv-L^{2\sigma-d} F_{d,\sigma}(tL^{\sigma})$. As
one sees equations (\ref{chiord}) and (\ref{finalchi}) are similar
and all what it has been written after equation (\ref{chiord})
remains true here. The conclusion we can draw from here is that
both {\it the ordinary perturbation theory} and the method based
on {\it the mode expansion} give {\it only one and the same}
result for the susceptibility in the region, where the bulk
properties of the system are dominating the critical behaviour of
the system i.e. in the region $L/\xi\gg1$.

\end{document}